\documentclass[10pt,emptycopyrightspace]{ewsn-proc}

\usepackage{graphicx}
\usepackage{balance}
\usepackage{comment}
\usepackage{subfig}
\usepackage[hidelinks]{hyperref}

\usepackage{hyperref}

\numberofauthors{1}

\author{
\alignauthor Merima Kulin, Eli de Poorter, Tarik Kazaz, Ingrid Moerman \\
    \affaddr{Department of Information Technology}\\
    \affaddr{Ghent University - imec, IDLab}\\
    \email{\{merima.kulin, eli.depoorter, tarik.kazaz, ingrid.moerman\}@intec.ugent.be}  
}

\title{Poster: Towards a cognitive MAC layer: Predicting the MAC-level performance in Dynamic WSN using Machine learning}

\begin{document}

\maketitle

\begin{abstract}
Predictable network performance is key in many low-power wireless sensor network applications.
In this paper, we use machine learning as an effective
technique for real-time characterization of the communication performance as observed by the MAC layer.
Our approach is data-driven and consists of three steps: extensive experiments for data collection, offline modeling and trace-driven performance evaluation. From our experiments and analysis, we find that a neural networks prediction model shows best performance.

\end{abstract}

\section{Introduction}
  \label{sec:intro}
  
Wireless sensor networks (WSN) have experienced explosive growth due to the promising and innovative application scenarios arising in the context of the Internet of Things (IoT). The availability of inexpensive low-power sensor devices has lead to an unprecedented surge in the number of connected devices. 
These devices generate traffic from heterogeneous radios that follow different medium access protocols and communication standards. A few examples in the 2.4GHz unlicensed band include IEEE 802.15.4, Bluetooth, WiFi, RFIDs, while in the sub-1GHz LoRA, SigFox, IEEE 802.11ah, etc.,  each suitable for a specific application domain. 
Such diversity of wireless technologies, applications and services will pose several communication challenges, including co-existence, cross-technology interference, spectrum scarcity and uncertainty in communication quality.

To tackle this problem, more sophisticated medium access control (MAC) protocols are needed. 
Traditional wireless MAC protocols are typically designed to optimally perform for low-power operation (e.g. ContikiMAC), low latency (e.g. CSMA/CA in low data rate), high throughput (e.g. 802.11 MAC) or high reliability (e.g. TSCH). 
\begin{figure}
\centering
\includegraphics[width=0.49\textwidth]{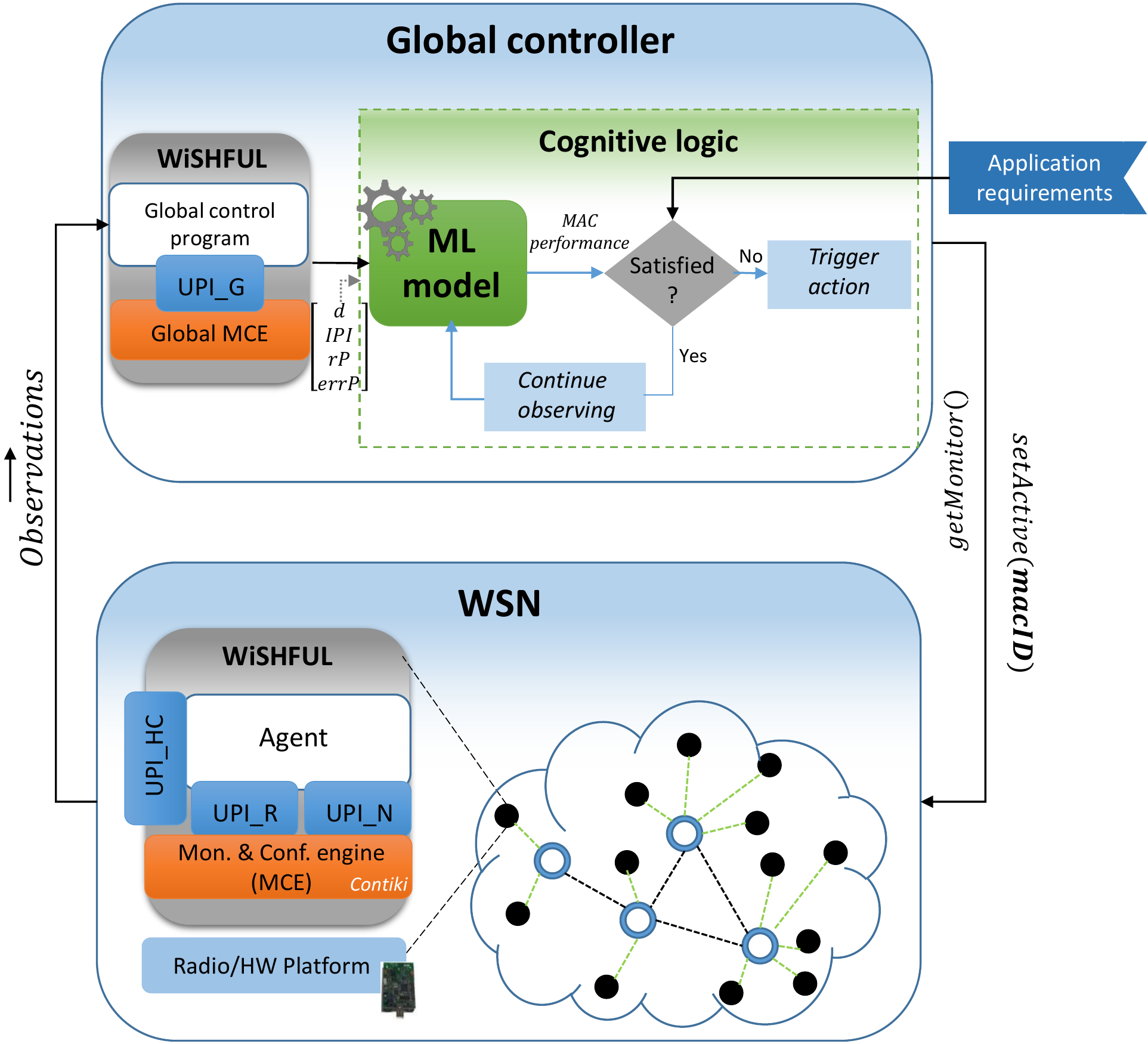}
\caption{System architecture for a cognitive MAC layer.}
\label{fig:cogmac}
\end{figure}
We argue that rather then developing optimized solutions for specific IoT application domains, a more suitable approach may be a \textit{cognitive MAC layer} that is able to dynamically choose the most appropriate MAC protocol in order to adapt to the observed network conditions and maintain the communication quality with regard to the application requirements. 

The idea of cognitive communication originates from the cognitive radio (CR) community, however, to this end their efforts have been focused on prototyping radios on top of powerful SDR platforms. 
In this work, we present a system design and first step implementation towards a cognitive MAC layer for constrained devices. We argue that the key component for an intelligent MAC layer is predictable performance. From a networking perspective, the most relevant aspect of reliable wireless communication quality is the packet delivery performance \cite{zhao2003understanding}.
Therefore, we use machine learning techniques to design a data-driven model \cite{kulin2016data} for packet delivery as a function of the collected measurements when running a CSMA/CA MAC protocol. Although an essential part for our cognitive MAC layer, the created machine learning model can act as a standalone reusable performance prediction component that can be integrated into other systems where performance prediction is necessary. 

\section{System design}

Figure \ref{fig:cogmac} presents the system architecture to accomplish a cognitive loop between two main components, the:

\textbf{\textit{Sensor network}} - a set of wireless nodes that generate information and are capable of reconfiguring its transmission parameters at runtime.

\textbf{\textit{Global controller}} - the central entity that collects and uses information from the wireless nodes to predict the MAC-level performance. Based on the predictions, it dynamically decides how to configure the MAC layer so as to improve the overall network performance (e.g. to cope with cross-technology interference it may decide to configure a more interference robust MAC protocol, e.g. TSCH). Finally, it disseminates the new configuration to the nodes.
At the heart of the global controller is a machine learning (ML) model that learns the environmental properties and uses the knowledge to predict the future MAC performance.
The following sections present the design of the ML model.

\textbf{Experimental setup}. To understand the MAC-level packet delivery performance we set up several experiments in the wilab2 testbed located in Ghent. We used 28 RM090 nodes with an IEEE 802.15.4 radio organized in a star-like network topology. All nodes use a CSMA/CA MAC protocol and periodically generate a $100B$ message to a single receiver located in the center of the topology. The transmission power is set to the maximum, i.e. $5dBm$, to ensure all nodes are in communication range.
To incorporate all factors that impact the MAC performance we set up several experiments varying the number of transmitting nodes (2-28 nodes), and the application traffic load (1 $pckt/2s$, 1 $pckt/s$, 2 $pckts/s$, 4 $pckts/s$, 8 $pckts/s$, 16 $pckts/s$ and 64 $pckts/s$). We used a USRP B210 to generate controllable interference patterns by transmitting a modulated carrier for 2 $ms$, followed by a 8 $ms$ idle period. More details and the data are available at \cite{merimakulin2017macperf}.

\textbf{Data collection.}
To simplify experiment control and data collection we created a global control program on the global controller using the Unified Programming Interfaces (UPIs) developed by the WiSHFUL project \cite{ruckebusch2016unified}. The UPIs provide the functionality to \textit{collect} measurements, \textit{configure} and on-the-fly \textit{adapt}  nodes that run a WiSHFUL compliant \textit{agent} program.
We run several experiments with the aforementioned setups and measured several aspects of the MAC-level performance, while the system was operating ($\sim$21$h$). 

\textbf{Feature space design.}
We extracted the most relevant features for predicting the MAC performance from consecutive observation intervals of the \textit{raw} data. Those are: number of detected nodes ($d$), inter-packet-interval ($IPI$), number of received packets ($rP$), number of erroneous packets/frames ($errP$).
Then we formed our feature vectors $ x^{(i)}=[d,IPI,rP,errP]^{T}$, and the corresponding communication reliability in terms of packet loss rate, $y^{(i)}=plr$.

\begin{figure}
\centering
\includegraphics[height=1.9in]{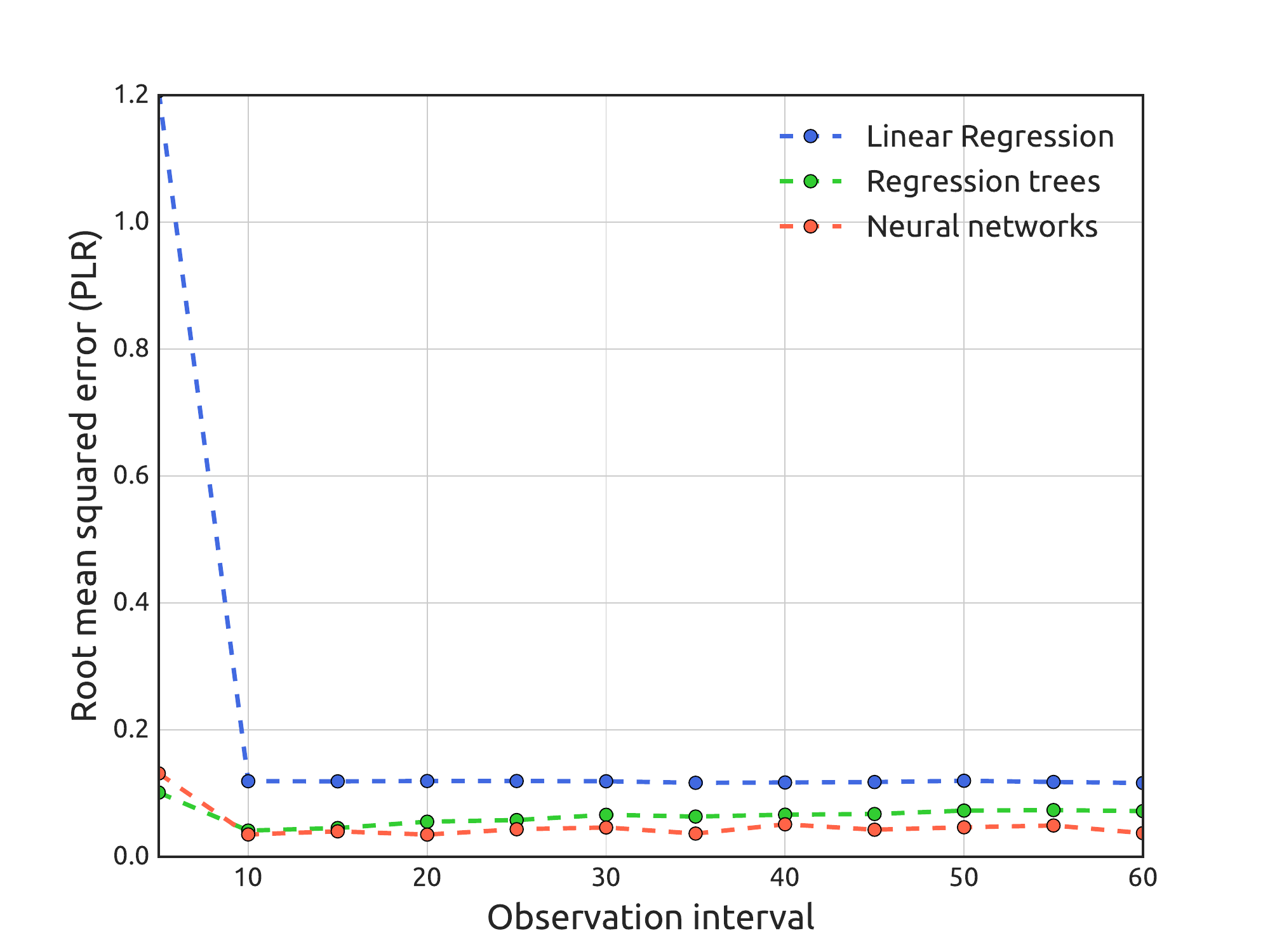}
\caption{RMSE of ML models vs. observation interval.}
\label{fig:rmse}
\end{figure}

\textbf{Machine learning models.}
Pairs of ($ x^{(i)}, y^{(i)}$) were used to train three machine learning algorithms: regression trees, linear regression and neural networks. We trained the algorithms for several observation intervals, and validated the performance of each model with a $10-fold$ cross-validation algorithm. Figure \ref{fig:rmse}
shows the root mean squared error (RMSE) for each model with tuned hyper-parameters. 
A neural network with 10 hidden layers ($HL$), 2000 training iterations, and a learning rate of $\alpha=0.1$, captured best the underlying distribution of the data. We selected the model for an observation interval of 30$s$ considering 30$s$ an appropriate interval for monitoring and deciding on the new MAC.

\begin{figure}
\centering
\includegraphics[height=1.9in]{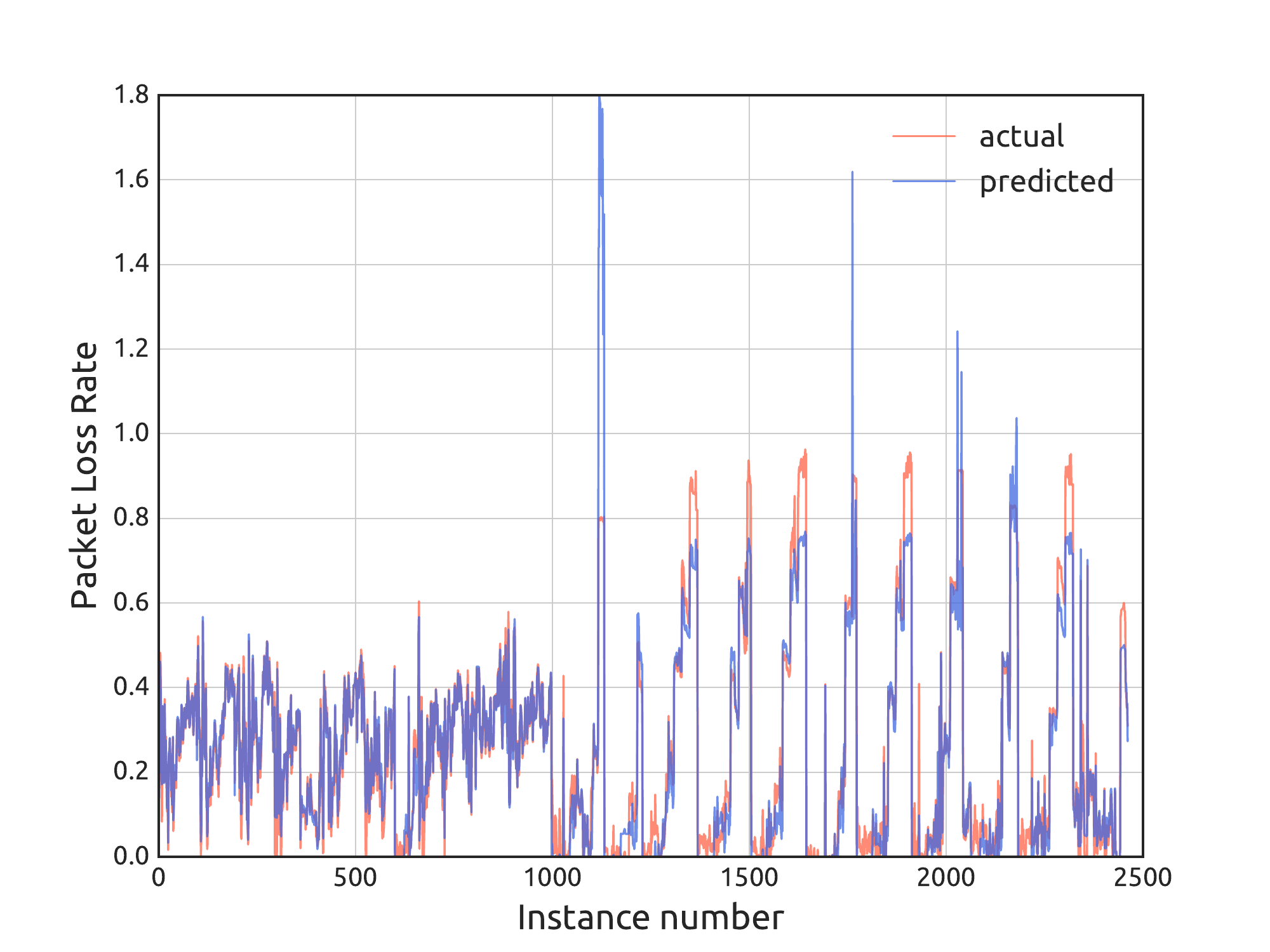}
\caption{Performance of neural networks with 10 $HL$}
\label{fig:test}
\end{figure}

\textbf{Performance evaluation.} To estimate how well the selected neural network model generalizes, we performed new experiments to collect new data ($\sim$21$h$) for testing. Figure \ref{fig:test} illustrates how well the model predicts $plr$ on instances from the test set.
It can be seen that the predictions are good in scenarios varying $d$ and $IPI$ (right-part) and also in scenarios with interference but a fixed medium IPI value (left-part).


\section{Acknowledgements}
The research leading to these results has received funding from the European Horizon 2020 Programmes under grant agreement n$^{\circ}$645274 
and n$^{\circ}$688116, and SBO SAMURAI.

%
%

\def\mybibdoicolor{\color{blue!75!black}} 
\newcommand*{\doi}[1]{\href{http://dx.doi.org/\detokenize{#1} {\raggedright\mybibdoicolor{DOI: \detokenize{#1}}}}}

%
%
\balance
\bibliographystyle{unsrt}
\nocite{*}
\bibliography{bibliogr}  
\end{document}